Enhancement of Spin Injection into Graphene by Water Dipping

K. M. McCreary,[1] Hua Wen,[1] H. Yu,[2] Wei Han,[1] E. Johnston-Halperin,[2] R. K. Kawakami[1,a)]

[1]*Department of Physics and Astronomy, University of California, Riverside, CA 92521*

[2]*Department of Physics, The Ohio State University, Columbus, OH 43210*

Abstract:

We immerse single layer graphene spin valves into purified water for a short duration (<1 min) and investigate the effect on spin transport. Following water immersion, we observe an enhancement in nonlocal magnetoresistance. Additionally, the enhancement of spin signal is correlated with an increase in junction resistance, which produces an increase in spin injection efficiency. This study provides a simple way to improve the signal magnitude and establishes the robustness of graphene spin valves to water exposure, which enables future studies involving chemical functionalization in aqueous solution.

[a)]Electronic mail: roland.kawakami@ucr.edu

Graphene has become a promising material for spin-based electronics due to the observation of spin transport at room temperature[1,2] with long spin lifetimes,[3,4] long spin diffusion lengths,[1] and high spin injection efficiency.[5] Additionally, its extreme sensitivity to surface adsorbates provides an effective method for modifying charge transport properties[6-8] and is predicted to induce interesting spin-dependent phenomena.[9,10] In particular, controlling graphene's properties using aqueous solutions for chemical functionalization is gaining interest in several fields.[11-14] To exploit such capabilities for spintronics, it is important to establish the robustness of graphene spin valves to aqueous solution processing. In this paper, we investigate the effect of dipping graphene spin valves into ultrapure water and surprisingly find that the spin transport signal is enhanced significantly following such a process. Further studies reveal that the enhancement of spin signal is correlated with an increase of the junction resistance, which can be understood within the framework of the one-dimensional (1D) drift-diffusion model[15] of spin transport.

Single layer graphene (SLG) spin valves, having cobalt electrodes with an MgO masking layer to reduce the Co/graphene contact area,[16] are fabricated following the procedure outlined in refs. 17, 18. Spin transport is investigated at room temperature in the nonlocal spin valve geometry[19] as summarized in figure 1(a). A representative curve for the nonlocal resistance, $R_{NL}$ =V/I, as a function of magnetic field, $H$, applied along the electrode axis is displayed in figure 1(b), with constant background subtracted. The nonlocal resistance difference between the parallel and antiparallel magnetization alignments of electrodes $E_2$ and $E_3$ is defined as $\Delta R_{NL}$. This represents the signal due to spin transport in the graphene from the spin injector ($E_2$) to the spin detector ($E_3$).

Prior to immersion in water, the graphene conductivity ($\sigma$) and $\Delta R_{NL}$ are characterized as a function of gate voltage ($V_G$), as shown in figure 2(a). The minimum in $\sigma$ identifies the Dirac point, $V_D$, at 5 V for sample 1. The electron density is given by $n = \alpha(V_G - V_D)$, with $\alpha = 7.2 \times 10^{10}$ cm$^{-2}$/V and negative values of $n$ corresponding to hole densities.[6] The spin signal, $\Delta R_{NL}$, is roughly proportional to $\sigma$, which can be understood by the 1D drift-diffusion model of spin transport.[15] In the limit of small but non-zero interface resistance, the spin signal is given by

$$\Delta R_{NL} = F_{el} F_{gr} \sigma$$
$$F_{el} = 4 \left( \frac{p_F R_F}{1 - p_F^2} + \frac{p_J R_J}{1 - p_J^2} \right)^2 \quad (1)$$
$$F_{gr} = \frac{W}{\lambda_s} \left( \frac{e^{(-L/\lambda_s)}}{1 - e^{(-2L/\lambda_s)}} \right)$$

where $W$ is the width of the graphene, $L$ is the spacing between injector ($E_2$) and detector ($E_3$), $\lambda_S$ is the spin diffusion length in the SLG, $R_J$ is the interfacial resistance between the Co and SLG, $p_J$ is the spin-asymmetry of the interfacial resistance, $R_F$ is the spin-resistance of the cobalt, and $p_F$ is the spin-asymmetry of the bulk Co resistivity. Equation 1 shows that $\Delta R_{NL}$ is proportional to $\sigma$, with the proportionality coefficient depending on the properties of the Co electrodes ($F_{el}$) and the spin diffusion in graphene ($F_{gr}$). This equation is valid in the limit of $R_J$, $R_F \ll \lambda_S/\sigma W$. Possible variations in $\lambda_S$ with electron density could produce a complicated relationship in the $\Delta R_{NL}$ vs. $V_G$ curve (through changes in $F_{gr}$), but the experimental results in figure 2(a) indicate this is not a strong effect.

To investigate the effect of water immersion on the spin transport properties, we submerge the spin valves in ultrapure water for approximately 5 seconds (unless otherwise noted) then quickly dry the devices under a flow of nitrogen gas. The water is prepared by filtration in a Millipore system, which yields a high resistivity of 18.2 MΩ-cm and a pH value between 5.70 and 5.85.

Figure 2(b) shows $\Delta R_{NL}$ as a function of gate voltage before and after the water dip for sample 1. Surprisingly, the water dipping enhances the spin signal by a factor of over 6. We compare σ before and after the water dip as shown in Figure 2(c) and find that although there are some minor changes in σ, it is clearly unable to explain the large enhancement of $\Delta R_{NL}$. In all devices showing an enhanced $\Delta R_{NL}$, the effect cannot be attributed to a corresponding enhancement of σ. This type of enhancement has been observed on 9 of 11 devices studied.

To investigate the origin of the enhanced spin signal, we turn our attention to the other factors in equation 1, namely $F_{gr}$ and $F_{el}$. $F_{gr}$ describes the spin diffusion from the injector electrode to the detector electrode with characteristic decay length of $\lambda_S$ ($\lambda_S = \sqrt{D\tau_S}$ where $D$ is the electron diffusion coefficient and $\tau_S$ is the spin lifetime). Even though the water dip does not substantially change the charge transport properties, namely σ and $D$, the spin diffusion length could increase if the water significantly increases $\tau_S$. As shown in figure 3(a) based on equation 1, when $R_J$ is held at 250 Ω, increasing $\lambda_S$ results in an enhancement of $\Delta R_{NL}$, which saturates as $\lambda_S$ becomes larger than $L$. All other parameters are specified in the figure captions. We note that further enhancement of $\Delta R_{NL}$ can occur if $\lambda_S$ exceeds the size of the graphene flake.[20, 21] Another factor in determining $\Delta R_{NL}$ is $F_{el}$, which is related to the electronic and spin-dependent properties of the electrodes and their interface with the graphene. The first term of $F_{el}$ in equation 1 depends on bulk properties of Co ($p_F$, $\lambda_S^{Co}$) that should not change with water dipping. Water dipping is more likely to affect the second term of $F_{el}$, which describes the interfacial properties of the junctions. The two important parameters are $p_J$ and $R_J$, which could be affected by the water dip. To illustrate the dependence of $\Delta R_{NL}$ on $p_J$ and $R_J$, figure 3(b) shows $\Delta R_{NL}$ as a function of $R_J$ while keeping $\lambda_S$ and $p_J$ fixed at typical values. The two curves correspond to $p_J$ values of 0.1 and 0.12. As apparent in the plot, an increase in $R_J$ can result in a substantial enhancement of $\Delta R_{NL}$.

In order to investigate the source of the observed magnetoresistance enhancement, we study several more devices in which close attention is paid to both the junction resistance and $\Delta R_{NL}$. To monitor the junction resistance, a three probe differential resistance (dV/dI) measurement is performed. In this geometry, the resistance of electrode $E_2$ is measured by applying a DC current plus AC modulation from $E_1$ to $E_2$ while measuring the voltage difference between $E_3$ and $E_2$ as the DC current is varied. Because the measured differential resistance $R_{el}$ includes not just $R_J$ but also the cobalt resistance, wire bond resistance, and the resistance of the measurement system we keep the wire bonds intact throughout the study including during the dip.

We find that the water dipping produces an increase in $R_{el}$ in a majority of samples. Furthermore, the enhancement of $\Delta R_{NL}$ was observed only in samples that also show an increase of $R_{el}$. The correlation between changes in $R_{el}$ and $\Delta R_{NL}$ is best illustrated in a particular device (sample 2) that was submerged in water twice. In a first dip, the $\Delta R_{NL}$ was not significantly affected, as can be seen by comparing the black (initial sample) and red (dip #1) curves in figure 4(a). The small increase in the $\Delta R_{NL}$ on the electron side of the Dirac point is attributed to the change in conductivity resulting from the water dip, as seen in figure 4(b). Additionally, the first dip had no effect on the electrode resistance, as indicated by the overlap of initial (solid black) and water dipped (dashed red) curves in figure 4(c). A second extended dip of ~50 sec was performed on the same sample. Following this dip, the $\Delta R_{NL}$ increased by more than 60% as shown in figure 4(a). The gate dependent conductivity curves before and after the extended dip are nearly identical, as can be seen in Figure 4(b), whereas the $R_{el}$ shows an increase from 1315 Ω to 1454 Ω. This sample highlights both situations we have observed as a result of exposure to water: either the $\Delta R_{NL}$ is enhanced and $R_{el}$ increases, or $\Delta R_{NL}$ remains largely unchanged and no change in $R_{el}$ is observed.

As discussed earlier, based on theoretical considerations, the possible explanations for the enhancement of $\Delta R_{NL}$ are increases in $\lambda_S$, $p_J$ and/or $R_J$. While changes to $\lambda_S$ and $p_J$ may be present (and certainly cannot be ruled out), the observed correlation between $R_{el}$ and $\Delta R_{NL}$ due to water dipping provides strong evidence that an increase of $R_J$ is the important factor for producing the enhanced spin signal. Furthermore, the relationship between increasing $R_J$ and $\Delta R_{NL}$ has been established experimentally in related studies on tunneling spin injection, where $R_J$ was intentionally increased by inserting tunnel barriers into the Co/graphene interface.[5] Further work is needed to understand the microscopic origin of the enhancement of $R_J$ due to water dipping.

In conclusion, we have demonstrated that dipping graphene spin valves in ultrapure water can enhance the nonlocal spin signal. Further studies provide evidence that an increase in the junction resistance is the most important factor for the enhancement of the spin signal, which can be understood within the 1D drift-diffusion model of spin transport. This could be useful in future device fabrication as a simple way to improve the signal magnitude. More importantly, it establishes the robustness of graphene spin valves to water dipping, which enables future studies of chemical functionalization in aqueous solution.

FIGURE CAPTIONS:

Figure 1: (a) A schematic illustration of a graphene spin valve in the nonlocal measurement geometry. (b) Typical nonlocal magnetoresistance curves as a function of applied magnetic field. This SLG sample has a channel length of 1.5 µm and is measured at the Dirac point. The black (red/grey) curve is for increasing (decreasing) magnetic field.

Figure 2: (a) The conductivity (solid line) and $\Delta R_{NL}$ (circles) as a function of gate voltage on sample 1 (with L= 2 µm), measured prior to water exposure. (b,c) $\Delta R_{NL}$ and $\sigma$ are compared before and after water immersion. $\Delta R_{NL}$ increases by at least a factor of six following water immersion, which cannot be attributed to the changes observed in conductivity.

Figure 3: Nonlocal magnetoresistance based on equation 1 using typical values of $W$=2 µm, $L$=1.5 µm, $\sigma$=0.5 mS, $p_F$=0.4, $R_F$=2.2×10$^{-2}$ Ω ($\rho_{Co}$=5.8×10$^{-8}$ Ωm, $\lambda_S^{Co}$=38 nm, $A_J$=0.1 µm$^2$). (a) The dependence of $\Delta R_{NL}$ on $\lambda_S$ while $R_J$ is held constant at 250 Ω and $p_J$ = 0.1. (b) The dependence of $\Delta R_{NL}$ on $R_J$ while $\lambda_S$ is held constant at 2 µm. The two curves are for values of $p_J$=0.1 and 0.12.

Figure 4: (a-c) $\Delta R_{NL}$, σ, and differential resistance $R_{el}$, respectively, before water immersion (black, solid) and after dip #1 (red, dashed) and dip #2 (blue, solid) for sample 2 having L=1.5 μm. The first dip has little effect on $\Delta R_{NL}$ and $R_{el}$, with the black and red $R_{el}$ curves completely overlapped. Following dip #2, the $\Delta R_{NL}$ in enhanced by at least 60%, and $R_{el}$ increases from 1315 Ω to 1454 Ω. The constant bias dependence of $R_{el}$ indicates an ohmic contact.

Acknowledgements:

We acknowledge technical assistance and discussion with A. Swartz, J. Wong, and M. J. Linman, and the support of NSF Grant No. MRSEC DMR-0820414, NSF Grant No. DMR-1007057, and ONR Grant No. N00014-09-1-0117.

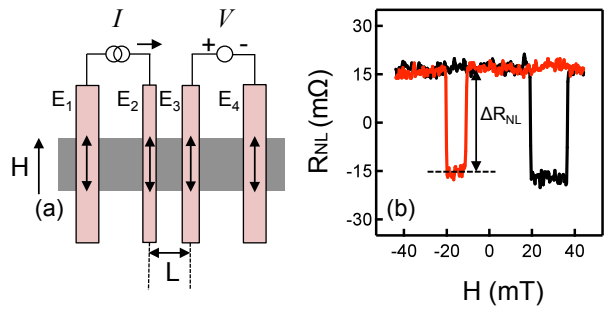

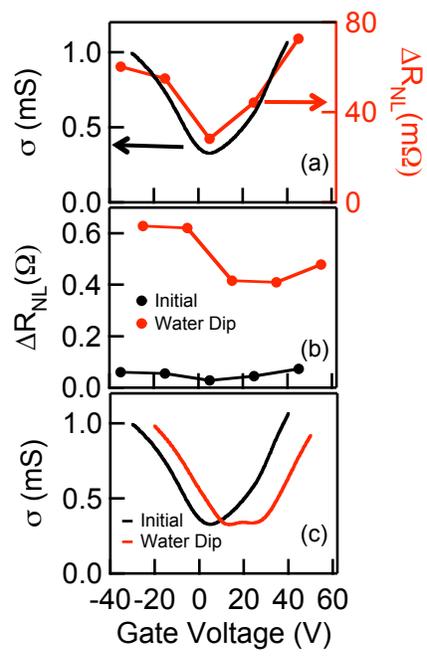

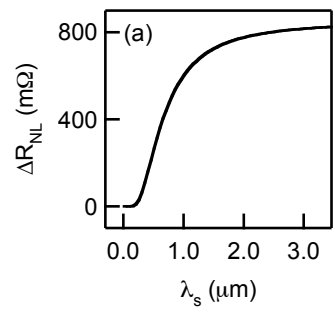 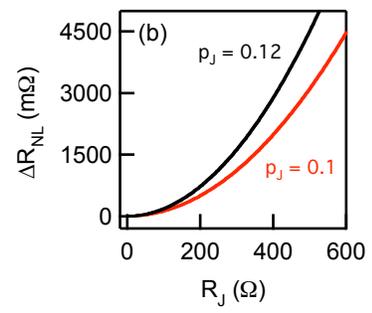

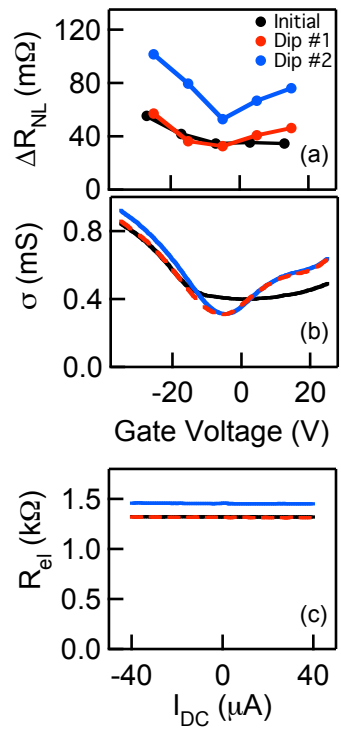